\documentstyle[preprint,aps]{revtex}
\begin{document}

\newcommand{\wt}{\widetilde}
\newcommand{\wh}{\widehat}

\title{ Effects of Impurity Vertex Correction on NMR \\
Coherence Peak in S-Wave Superconductors}
\author{Han-Yong Choi}
\address{Department of Physics, Sung Kyun Kwan University,
Suwon 440-746, Korea}
\author{E. J. Mele}
\address{Department of Physics, University of Pennsylvania,
Philadelphia, PA 19104, USA}
\date{Feb. 10, 1995}
\maketitle
\begin{abstract}
We study the effects of non-magnetic impurity vertex correction on nuclear
spin-lattice relaxation rate  $1/{T_1} $ of
conventional s-wave superconductors within the Eliashberg formalism.
We obtain, with a self-consistent $t-$matrix treatment of impurity
scatterings,
the expressions for impurity vertex function and nuclear spin-lattice
relaxation rate.
The $1/T_1$ is evaluated with a simple approximation on angular average,
and found to agree in clean limit with the previous result
that $1/(T_1 T)$ remains unrenormalized under the impurity vertex correction.
As dirtiness is increased, on the other hand,
the coherence peak in $1/(T_1 T)$ is found to
increase due to the impurity vertex correction.
This result is discussed in connection with
the experimental observations on conventional superconductors.

\end{abstract}
\pacs{PACS numbers: 74.20.Fg, 74.25.Nf, 74.70.Ad}

Nuclear magnetic resonance (NMR) relaxation rate $T_1 ^{-1}$
measurements have been used extensively to study both normal
and superconducting properties of various materials \cite{slichter}.
In simple metals, the
conduction electrons provide the dominant relaxation channel for polarized
nuclear spins, which results in the well-known Korringa law
$T_1 ^{-1} \propto ~T$. Below the transition temperature, $T_c$,
the superconducting gap opens and, consequently, density of states (DOS)
increases. The former and the latter, respectively, suppresses and increases
the relaxation rate, so that the relaxation rate is determined by the
competition between the two: At temperatures immediately below
$T_c$, the DOS effect dominates and there is an increase of ${(T_1 T)}^{-1}$
relative to the normal state Korringa value at $T = T_c$.
At the low temperature regime, on the other hand,
the effect of
gap opening takes over and the relaxation rate falls off exponentially
or algebraically depending on the symmetry of the superconducting gap.
This peak as a function of temperature is known as
coherence peak, or Hebel-Slichter peak in NMR, whose observation
was very important in establishing the Bardeen-Cooper-Schrieffer (BCS)
theory \cite{bcs,schrieffer,allen2}.

There are several factors that affect the  NMR
coherence peak of ${(T_1 T)}^{-1}$ \cite{nmrrev}.
The suppression of NMR coherence peak may be attributed to
(a) gap anisotropy \cite{nmrrev}/non s-wave pairing
of superconducting phase \cite{pines,scalapino},
(b) strong coupling damping \cite{allen1,nakamura,carbotte},
(c) paramagnetic impurities in samples \cite{nmrrev,maki1}, and/or
(d) strong Coulomb interaction such as paramagnon/antiparamagnon effects
\cite {hasegawa,monien}.
The non-magnetic impurity scatterings, on the other hand, have no influence
on ${(T_1 T)}^{-1}$ \cite{maki1,agd,anderson}, other than the smearing of
gap anisotropy.
This is because (a) there exists a simple
scaling relation between the renormalization function of pure ($Z(\omega)$)
and impure ($\wt{Z} (\omega)$) superconductors and (b) in the expression for
${(T_1 T)}^{-1}$,
the numerator and denominator are of the same powers in
$\wt{Z}(\omega)$, as will be discussed in more detail later.
Experimentally, it was observed
in a series of NMR experiments on Al- \cite{masuda} and In-based alloys
\cite{maclau}
that the coherence peak in ${(T_1 T)}^{-1}$ was
increased as the samples become dirtier, or the scattering lifetime,
$\tau$, is decreased.
This is understood to be due to the increase of DOS as mentioned above
as a result of gap anisotropy smearing by
impurity scatterings \cite{nmrrev,clem}. We point out in this paper that
this may also be understood in terms of impurity vertex correction (IVC).
We found, as will be detailed below, that in the weak coupling limit
the ratio of the coherence peak to normal state Korringar value roughly doubles
as one goes from clean to dirty limit in rough agreement with experimental
observations \cite{nmrrev,masuda,maclau}.
The enormous enhancement of NMR coherence peak reported in some
organic superconductors, on the other hand, may be attributed to
thermal fluctuations of magnetic flux lines \cite{flux}.

It is well known that the IVC is very important
in the transport properties of metals \cite{mahan}.
The scattering lifetime $\tau$ is
changed to the transport lifetime $\tau_{tr}$ due to IVC, which appears in,
for example, $\sigma (0) = ne^2 \tau_{tr} /m$, where $n, ~e$ and $m$ are,
respectively, the number density, charge and effective mass of the charge
carriers.
Because $T_1^{-1}$ and microwave conductivity,
$\lim_{\omega \rightarrow 0} \sigma(\omega)$, have the same ``plus coherence
factor'' \cite{maki1}, we may expect that the IVC may also be important for
NMR relaxation rate. Maki and Fulde (MF)
\cite{maki2} studied this problem some 30 years ago assuming an infinite Fermi
energy, $\epsilon_F$. They found the
impurity vertex function $\wh{ \Gamma} (\omega, \vec{q})
\rightarrow 1$ in the
limit of the momentum transfer $\vec{q} \rightarrow \infty$ and $\omega
\rightarrow 0$, and concluded that $T_1^{-1}$, consequently,
remains unrenormalized under IVC.
This conclusion, however, is not rigorous because $T_1^{-1}$
is given by an integral over the momentum transfer $\vec{q}$, not by the
limit $\vec{q} \rightarrow \infty$.
As will be discussed later, $T_1^{-1}$ involves integrals over $\vec{k'}$
and $\vec{q}$ of the product $Tr \Bigl[ \wh{G}(\vec{k'}) \wh{G}
(\vec{k'}+\vec{q}) \wh{\Gamma}(\vec{q}) \Bigr] $, where $Tr$ and $G$
stand for trace and Green's function,
respectively. Since superconducting pairing
occurs mainly within a narrow region around the Fermi surface, we can
see that the main contribution to $T_1^{-1}$ comes from the region
$|\vec{k'}| \approx p_F$ and $|\vec{k'}+\vec{q}| \approx p_F$,
that is, from the region $Q = (q^2 +2\vec{q} \cdot \vec{k'})/2m \approx 0$,
where $p_F$ is the Fermi momentum.
The $Q \approx 0$ region and the $q \rightarrow \infty$ considered by
MF should be distinguished.
Moreover, the impurity vertex function which enters $T_1^{-1}$
has a singularity as $q \rightarrow 0$, which contributes to
the $Q \approx 0$ region, as will be discussed.
These observations call for a more careful analysis of
the IVC effects on spin-lattice relaxation rate
than the $q \rightarrow \infty$
of MF. In this paper, we report such a calculation.
We find an explicit expression for $T_1^{-1}$ including IVC
with a simple approximation about angular average which
agrees with the approximate conclusion
of MF that $T_1^{-1}$ remains unrenormalized under IVC
for $\ell/\xi \gg 1$,
where $\ell = v_F \tau$ is the electron mean free path
and $\xi = \hbar v_F / \pi \Delta_0$ is the superconducting
coherence length, $\Delta_0$ the gap function at zero temperature.
As $\ell/\xi$ is reduced, the NMR coherence peak is found to increase
due to IVC.

The nuclear spin-lattice relaxation rate
$T_1 ^{-1}$ is given by \cite{moriya}
\begin{eqnarray}
\frac{1}{T_1} = \lim_{\omega \rightarrow 0}
\frac{1}{1 - e^{-\beta \omega} } \sum_{\vec{q}} \Big| A_q \Big|^2
Im \Bigl[ \chi_{+-} (\omega + i \delta ,\vec{q}) \Bigr],
\label{t1}
\end{eqnarray}
where $A_q$ is a form factor related with the conduction electron
wavefunctions, and $\chi_{+-} ( \omega , \vec{q}) $ is
a spin-spin correlation function at frequency $\omega$ and momentum transfer
$\vec{q}$, obtained through analytic continuation from the Matsubara function
$\chi_{+-} ( i\omega , \vec{q}) $.
It should be calculated with non-magnetic
impurities included fully self-consistently. We do this by renormalizing
the quasi-particle Green's function due to impurity scattering in the
$t$-matrix
approximation and also by including the impurity vertex correction in ladder
approximation, in accordance with the Ward identity \cite{schrieffer,mahan}.
This procedure results in the Bethe-Salpeter equation,
which is to be solved for the vertex function.
The renormalized $2 \times 2$ matrix single particle Matsubara Green's
function due to impurity scattering is given by
\begin{eqnarray}
\wh{G}(ip_n ,\vec{k} ) = - \frac{i\wt{W}_n +
\wt{\xi}_k \tau_3
+\wt{\phi}_n \tau_1 }
{ {\wt{W}_n}^2 +{\wt{\xi}_k }^2 +{\wt{\phi}_n }^2},~~
\label{green}
\end{eqnarray}
where $\tau_i$'s are the Pauli matrices operating on the Nambu space.
$i\wt{W}_n = ip_n \wt{Z}_n$, $\wt{\phi}_n =
\wt{\Delta}_n \wt{Z}_n$,
where $p_n = \pi (2n+1) /\beta$ is the Matsubara frequency, and
$\wt{Z}_n =\wt{Z}(ip_n) $ and
$\wt{\Delta}_n = \wt{\Delta}(ip_n)$ are,
when analytically continued to real frequency,
the renormalization function and gap function, respectively.
A tilde on a variable denotes that it is renormalized by impurity scatterings,
and $\xi_k = {k^2}/{2m} -\mu$, where $\mu$ is the chemical
potential. The self-energy, $\wh{\Sigma} (ip_n , \vec{k}) $,
due to impurity scatterings is then given by
\begin{equation}
\wh{G}^{-1} (ip_n , \vec{k}) =
{\wh{G}_0}^{-1}(ip_n , \vec{k}) - \wh{\Sigma}(ip_n , \vec{k}),~~
\wh{\Sigma}(ip_n , \vec{k}) = n_i \wh{T}(ip_n , \vec{k}),
\label{g-dyson}
\end{equation}
as shown in Fig. 1(a),
where $n_i$ is impurity concentration, $\wh{T}$ the $t$-matrix
for impurity scattering, and $\wh{G}_0$ is the Green's function for
pure superconductors.
We treat the non-magnetic impurity scattering in a self-consistent
$t$-matrix approximation as shown in Fig. 1(b) to obtain
\begin{equation}
\wh{T} = \frac{1}{\pi N_F} \frac{U}{1+U^2} \biggl[ \tau_3
- \frac{U}{\sqrt{\wt{W}_n^2 +\wt{\phi}_n^2} }
\biggl( i\wt{W}_n -\wt{\phi}_n \tau_1
\biggl) \biggl].
\label{t-mat}
\end{equation}
Then there exists a simple scaling relation between pure ($iW_n,~\phi_n$)
and impure ($i\wt{W}_n, ~\wt{\phi}_n$) superconductivity
\cite {maki1,agd}.
\begin{eqnarray}
i\wt{W}_n &=& \eta_n iW_n, ~~ \wt{\phi}_n = \eta_n \phi_n,~~
\wt{\xi} = \xi + \frac{1}{2\tau U},
\nonumber \\
\eta_n &=& 1 + \frac{1} {2\tau \sqrt{W_n^2 +\phi_n^2} }, ~~
\frac{1}{2\tau} = \frac{n_i}{\pi N_F} \frac{U^2}{1+U^2}
= \frac{n_i}{\pi N_F} \sin^2{\delta_N},
\label{scale}
\end{eqnarray}
where $n_i$ is impurity concentration and $\delta_N$ the phase shift
due to impurity scattering, $U = \pi N_F V = \tan{\delta_N}$, where
$V$ is the impurity scattering potential energy.

The impurity vertex function, $\wh{\Gamma}$, is calculated in
ladder approximation as shown in Fig. 1(c). We have
\begin{equation}
\wh{\Gamma} (ip_n,ip_m,\vec{q}) = \tau_0 + \sum_{\vec{k}'}
\wh{T}(ip_n) \wh{G}(ip_n,\vec{k}')
\wh{\Gamma} (ip_n,ip_m,\vec{q})
\wh{G}(ip_m,\vec{k}'+\vec{q}) \wh{T}(ip_m ).
\label{vert}
\end{equation}
This is the Bethe-Salpeter equation at arbitrary momentum transfer
$\vec{q}$ and frequency $i\omega_{m-n} = ip_m -ip_n = 2\pi (m-n)/\beta$
in ladder approximation,
which we need to solve for $\wh{\Gamma}$.
Then $T_1^{-1}$ can be calculated from Eq.\ (\ref{t1}) and
\begin{eqnarray}
\chi_{+-}(i\omega) &=& \frac{1}{\beta} \sum_{ip} P(ip,ip+i\omega),
\nonumber \\
P(ip,ip+i\omega) &=&
\sum_{\vec{k'}, \vec{q} } Tr \Bigl[
\wh{G}(ip+i\omega,\vec{k}'+\vec{q})
\wh{G}(ip,\vec{k}') \wh{\Gamma}
(ip,ip+i\omega,\vec{q}) \Bigr].
\label{chi2}
\end{eqnarray}
We will omit the subscripts of the Matsubara frequency for convenience.
After the analytic continuation of $i\omega
\rightarrow \omega + i\delta$ and the limit $\omega \rightarrow 0$, Eq.\
(\ref{chi2}) is reduced to \cite{mahan}
\begin{equation}
\lim_{\omega \rightarrow 0} ~\frac{1}{\omega}
Im \Bigl[ \chi_{+-} (\omega +i\delta) \Bigr]
= \int_{-\infty}^{\infty} \frac{d\epsilon}{2\pi} ~
\frac{\partial f(\epsilon)}{\partial \epsilon} \biggl\{
P(\epsilon -i\delta,\epsilon+i\delta) -Re \Bigl[
P(\epsilon +i\delta,\epsilon+i\delta) \Bigr] \biggr\},
\label{analy}
\end{equation}
where $f(\epsilon) = 1/(1+e^{\beta \epsilon})$ is the Fermi distribution
function.
In order to solve Eq.\ (\ref{vert}), we employ the following factorization
as with Hirschfeld $\it{et.~ al.}$ \cite{hirsch}.
\begin{eqnarray}
L_{ij} (ip,ip+i\omega,\vec{q}) &=& \frac{1}{2} \sum_{\vec{k'}}
Tr \Bigl[ \tau_i \wh{ G}(ip,\vec{k}') \tau_j
\wh{ G}(ip + i\omega,\vec{k}'+\vec{q}) \Bigr],
\nonumber \\
S_{ij} (ip,ip+i\omega) &=&
\frac{1}{2} Tr \Bigl[ \tau_i \wh{T}(ip)
\tau_j \wh{T} (ip+i\omega) \Bigr],
\nonumber \\
\Gamma_i (ip,ip+i\omega,\vec{q}) &=& \frac{1}{2} Tr
\Bigl[ \tau_i \wh{\Gamma} \Bigr].
\label{lij}
\end{eqnarray}
Then Eq.\ (\ref{vert}) is reduced to
\begin{equation}
\Gamma_i = \delta_{i,0} + \sum_{k,l} S_{ik} L_{kl} \Gamma_l ,
\label{vert3}
\end{equation}
and Eq.\ (\ref{chi2}) to
\begin{equation}
\chi_{+-}(i\omega) = \sum_{ip, \vec{q} }
\sum_j L_{0j} \Gamma_j .
\label{chi3}
\end{equation}
The summations over $\vec{k'}$ in
Eq.\ (\ref{chi2}) are carried out with a constant density of states,
$\sum_{\vec{k'}} = \int_{-\infty}^{\infty} N_F d\xi_{k'} ~
\frac{1}{2}\int_{-1}^{1} d\nu$,
where $\nu = \cos(\theta)$, $\theta$ is the angle between $\vec{k'}$
and $\vec{q}$, and with the approximation
$\xi_{\vec{k'}+\vec{q}} =
{(\vec{k'}+\vec{q})}^2 /2m -\mu \approx \xi_{\vec{k'}} + Q(\nu ), $
where
$Q(\nu) = (2p_F q \nu +q^2)/2m$,
which are standard approximations in
the theory of superconductivity \cite {maki1}.
It is then straightforward to evaluate $L_{ij}$. For instance,
\begin{eqnarray}
L_{00} (ip,ip+i\omega,q) &=& N_F
\frac{\pi \tau}{2} \bigl<f_{-} (ip,ip+i\omega,q) \bigr>
\bigl( - \phi_n \phi_m +W_n W_m -\sqrt{}_n \sqrt{}_m \bigr)
/\sqrt{}_n \sqrt{}_m ,
\label{l00} \\
\bigl<f_{\pm} (ip,ip+i\omega,q) \bigr> &=& \frac{2i}{\tau} \biggl[
\biggl< \frac{1} {Q -i{\sqrt{\wt{} }}_n -
i{\sqrt{\wt{} }}_m } \biggr> \pm
\biggl< \frac{1} {Q +i{\sqrt{\wt{} }}_n +
i{\sqrt{\wt{} }}_m } \biggr> \biggr]
\nonumber \\
&=& \frac{i}{4\tau \epsilon_F x} \biggl[ \log \biggl(
\frac{x^2 +x -i\delta_s}{x^2 -x -i\delta_s} \biggr)
\pm \log \biggl( \frac{x^2 +x +i\delta_s}
{x^2 -x +i\delta_s} \biggr) \biggr] ,
\label{fpm}
\end{eqnarray}
where $\sqrt{}_n = \sqrt{W_n^2 +\phi_n^2}$,
${\sqrt{\wt{} }}_n = \sqrt{\wt{W}_n^2 +{\wt{\phi}}_n^2 }$,
$x = q/(2p_F)$, and
$\delta_s = ({\sqrt{\wt{} }}_n +{\sqrt{\wt{} }}_m)/\epsilon_F
=(\sqrt{}_n +\sqrt{}_m +1/\tau)/\epsilon_F$
using the scaling relation of Eq.\ (\ref{scale}).
$\bigl<~ \bigr>$ represents an angular average over $\nu$,
that is, $ \bigl<F \bigr> = \textstyle{\frac{1}{2}
\int_{-1}^{1} d\nu F(\nu) }$.
Note that Eq.\ (\ref{l00}) is expressed in terms of untilded variables
taking advantage of Eq.\ (\ref{scale}) because its numerator and denominator
are of the same powers in $\wt{W}$ and $\wt{\phi}$.

Let us first consider the Born limit of $\delta_N \rightarrow 0$.
Then, $\wh{T} = V \tau_3$, $\frac{1}{2\tau} = n_i \pi N_F V^2$,
and $S_{ij}$ of Eq.\ (\ref{lij}) becomes a diagonal matrix.
In this case we can find an explicit solution
$\Gamma(i\omega, \vec{q}) $ to Eq.\
(\ref{vert3}) for arbitrary $\vec{q}$ and $i\omega$.
We will discuss the general case of finite $\delta_N$ later.
After somewhat laborious calculations, we obtain
\begin{eqnarray}
\Gamma_0 &=& 1 +
\frac{\bigl<f_- \bigr> +\bigl<f_-\bigr>^2 -\bigl<f_+\bigr>^2 }
{(1 +\bigl<f_-\bigr> +\bigl<f_+\bigr>) (1 +\bigl<f_-\bigr> -\bigl<f_+\bigr>)}
\frac{W_n W_m -\phi_n \phi_m -\sqrt{}_n \sqrt{}_m }{2\sqrt{}_n \sqrt{}_m },
\nonumber \\
\Gamma_1 &=&
\frac{\bigl<f_- \bigr> +\bigl<f_-\bigr>^2 -\bigl<f_+\bigr>^2 }
{(1 +\bigl<f_-\bigr> +\bigl<f_+\bigr>) (1 +\bigl<f_-\bigr> -\bigl<f_+\bigr>)}
\frac{iW_n \phi_m +iW_m \phi_n}{2\sqrt{}_n \sqrt{}_m },
\nonumber \\
\Gamma_2 &=&
\frac{\bigl<f_+\bigr> }{(1 +\bigl<f_-\bigr> +\bigl<f_+\bigr>)
(1 +\bigl<f_-\bigr> -\bigl<f_+\bigr>)}
\frac{W_n \sqrt{}_m -W_m \sqrt{}_n }{2\sqrt{}_n \sqrt{}_m },
\nonumber \\
\Gamma_3 &=&
\frac{-\bigl<f_+\bigr> }{(1 +\bigl<f_-\bigr> +\bigl<f_+\bigr>)
(1 +\bigl<f_-\bigr> -\bigl<f_+\bigr>)}
\frac{\phi_n i\sqrt{}_m +\phi_m i\sqrt{}_n }{2\sqrt{}_n \sqrt{}_m }.
\label{gamma}
\end{eqnarray}
We substitute this into Eq.\ (\ref{chi3}) to finally obtain
\begin{eqnarray}
\chi_{+-}(i\omega) &=& \sum_{ip }
\biggl[ \frac{-\phi_n \phi_m +W_n W_m}{ \sqrt{}_n \sqrt{}_m} -1 \biggr ]
{}~\Lambda (ip,ip+i\omega),
\nonumber \\
\Lambda (ip,ip+i\omega) &=& \int_0^{\infty} dq q^2
\frac{ \bigl<f_-\bigr> +\bigl<f_-\bigr>^2 +\bigl<f_+\bigr>^2
(W_n W_m +\phi_n \phi_m )/(2\sqrt{}_n \sqrt{}_m ) }
{(1 +\bigl<f_-\bigr> +\bigl<f_+\bigr>)
(1 +\bigl<f_-\bigr> -\bigl<f_+\bigr>)} .
\label{renor}
\end{eqnarray}
Note that in the limit $q$ (or $x$) $\rightarrow 0$,
$\bigl< f_- (\epsilon-i\delta,\epsilon+i\delta) \bigr>$ and
$ \bigl< f_+ (\epsilon-i\delta,\epsilon+i\delta) \bigr>$
of Eq.\ (\ref{fpm}), which we need for analytic continuation as in Eq.\
(\ref{analy}),
reduce to $\--1 +\textstyle{\frac{1}{3}} (\epsilon_F \tau x)^2
+{\cal O} (x^4 ) $ and $i \epsilon_F \tau x^2 +{\cal O} (x^4 ) $, respectively.
The vertex functions given by Eq.\ (\ref{gamma}), consequently, diverge as
$q \rightarrow 0$, which is canceled
by the phase factor of $q^2$ in Eq.\ (\ref{renor}).
We need to carry out the summation over the Matsubara frequency $ip$
and the momentum $\vec{q}$ in Eq.\ (\ref{renor}), and
then take analytic continuation and the limit
$\omega \rightarrow 0$ as given in Eq.\ (\ref{analy}).
Neglecting IVC amounts to taking $\Lambda$ = 1, for which we obtain
the well known result \cite{fibich}
\begin{eqnarray}
\frac{1}{T_1 T} &\propto& \int_0^{\infty} d\epsilon
\frac{\partial f}{\partial \epsilon}~
\biggl\{ \biggl[ Re \biggl(
\frac{\epsilon}{\sqrt{\epsilon^2 -\Delta^2}} \biggr) \biggr]^2
+ \biggl[ Re \biggl(
\frac{\Delta}{\sqrt{\epsilon^2 -\Delta^2}} \biggr) \biggr]^2
\biggr\}.
\label{t1-s} \\
&\rightarrow & \int_{\Delta}^{\infty} d\epsilon
\frac{\partial f}{\partial \epsilon}~ \biggl\{
\frac{\epsilon^2 +\Delta^2}{\epsilon^2 -\Delta^2} \biggr\}.
\label{t1-w}
\end{eqnarray}
Eq.\ (\ref{t1-w}) follows in the weak coupling limit.
Eq.\ (\ref{renor}) may not be evaluated analytically, but
is greatly simplified in the following
approximation that treats the angular average differently:
\begin{eqnarray}
\Lambda \approx \int_0^{\infty} dq~ q^2 ~ \biggl<
\frac{ f_- + f_-^2 +f_+^2 (W_n W_m +\phi_n \phi_m )/(2\sqrt{}_n \sqrt{}_m ) }
{ (1 +f_- +f_+ ) (1 +f_- -f_+)} \biggr> .
\label{angular}
\end{eqnarray}

With the approximation of Eq.\ (\ref{angular}),
we find, in the weak coupling limit, after the analytic continuation
of Eq.\ (\ref{analy})
\begin{equation}
\frac{1}{T_1 T} \propto \int_{\Delta}^{\infty} d\epsilon
\frac{\partial f}{\partial \epsilon}~ \biggl\{
\frac{\epsilon^2}{\epsilon^2 -\Delta^2} +
\frac{\Delta^2}{\epsilon^2 -\Delta^2}
\biggl[ 1 + \frac{2}{1+(4 \tau )^2 (\epsilon^2 -\Delta^2 )}
\biggr] \biggl\}.
\label{t1t-f}
\end{equation}
We also considered the case where the impurity potential is not weak, that is,
where the phase shift $0 < \delta_N \leq \pi/2$. The resulting equation is
extremely complicated. For $\omega \rightarrow 0$, however, we were able to
find an explicit expression corresponding to Eq.\ (\ref{t1t-f})
with the approximation of Eq.\ (\ref{angular}): Eq.\
(\ref{t1t-f}) is still valid for arbitrary phase shift. All that is needed
is that $\frac{1}{2\tau} = n_i \pi N_F V^2$ of the Born limit should be
extended to $\frac{1}{2\tau} = \frac{n_i}{\pi N_F} \frac{U^2}{1+U^2}$ as
given in Eq.\ (\ref{scale}).
Eq.\ (\ref{t1t-f}) reduces to Eq.\ (\ref{t1-w})
in the limit $\tau \rightarrow \infty$.
This agrees with the previous result of MF that $1/T_1$ remains
unrenormalized under the IVC.
In the limit $\tau \rightarrow 0$, on the other hand,
\begin{equation}
\frac{1}{T_1 T}
\rightarrow  \int_{\Delta}^{\infty} d\epsilon
\frac{\partial f}{\partial \epsilon}~ \biggl\{
\frac{\epsilon^2 + 3\Delta^2}{\epsilon^2 -\Delta^2} \biggr\}.
\label{t1-dirty}
\end{equation}

The dominant contribution to $1/(T_1 T)$ comes from the region
$\epsilon \approx \Delta$. Consequently, comparing
Eqs.\ (\ref{t1-w}) and (\ref{t1-dirty}), we can easily see that
the ratio of coherence peak to normal state value is bigger in
dirty limit than in clean limit by a factor of about 2.
In Fig. 2, we show the normalized NMR relaxation rate,
$(T_1 T)_n /(T_1 T)_s$, calculated from Eq.\ (\ref{t1t-f}),
as a function of temperature for several
values of $\ell / \xi$. We took the electron-phonon coupling constant
$\lambda = 0.5$ and the phonon frequency $ \omega_{ph} = 0.1$ eV for
representative values.
The solid, dot-dot-dashed, dot-dashed, and dashed curves correspond to
$\ell /\xi$ = 100, 10, 1, and 0.1, respectively.
The curves of $\ell/\xi$ = 0.1 and 100 are, respectively, almost
indistinguishable from those of the dirty and clean limit.
The NMR coherence peak is increased by a factor of 1.7 as expected, as
$\ell/\xi$ is decreased from 100 to 0.1,
which is in fair agreement with the experimental observations on Al- and
In-based alloys \cite{nmrrev,masuda,maclau}.
This observation raises an interesting possibility that the experimental
observation of the NMR coherence peak increase may be understood
in terms of impurity vertex correction rather than the gap anisotropy
smearing. Experimental distinction between the two effects is highly
desirable in this regard.

To summarize, we considered the effects of impurity vertex correction on
nuclear spin-lattice relaxation rate within the Eliashberg formalism.
The relaxation rate was evaluated
and found to agree in clean limit with the previous approximate conclusion
of Maki and Fulde that $1/(T_1 T)$ remains unrenormalized under the impurity
vertex correction.
As the scattering lifetime is decreased, on the other hand,
the coherence peak in $1/(T_1 T)$ was found to
increase due to the impurity vertex correction.
The nuclear spin-lattice relaxation data
measured on conventional superconductors
as impurity scattering rates are varied may also
be understood in terms of impurity vertex correction rather than
the gap anisotropy smearing effects induced by impurity scatterings.

H.Y.C. was supported by Korea Science and Engineering Foundation
through Grant No. 931-0200-003-2 and through
Center for Theoretical Physics, Seoul National University,
and by the Ministry of Education through Grant No. BSRI-94-2428.

\begin{figure}
{\bf FIG. 1}.
Feynmann diagrams for (a) self-energy, $\wh{\Sigma}$,
(b) {\it t-}matrix for impufity scatterings, $\wh{T}$,
and (c) impurity vertex function, $\wh{\Gamma}$.
The solid double and single lines, respectively, stand for
the renormalized, $\wh{G}$, and bare Green's functions,
$\wh{ G}_0$.
The dashed double and single lines, respectively, stand for
the {\it t-}matrix, $\wh {T}$,
and bare impurity scatterings, $V \tau_3$, and the shaded
triagle in (c) represents the impurity vertex function, $\wh{\Gamma}$.
\vspace{0.5in}

{\bf FIG. 2}.
The normalized NMR relaxation rate, $(T_1 T)_n /(T_1 T)_s$, as a function
of $T/T_c$. The solid, dot-dot-dashed, dot-dashed, and dashed lines were,
respectively, calculated for $\ell/\xi$ = 100, 10, 1, and 0.1
with $\lambda$ = 0.5 and $\omega_{ph}$ = 0.1 eV.

\end{figure}

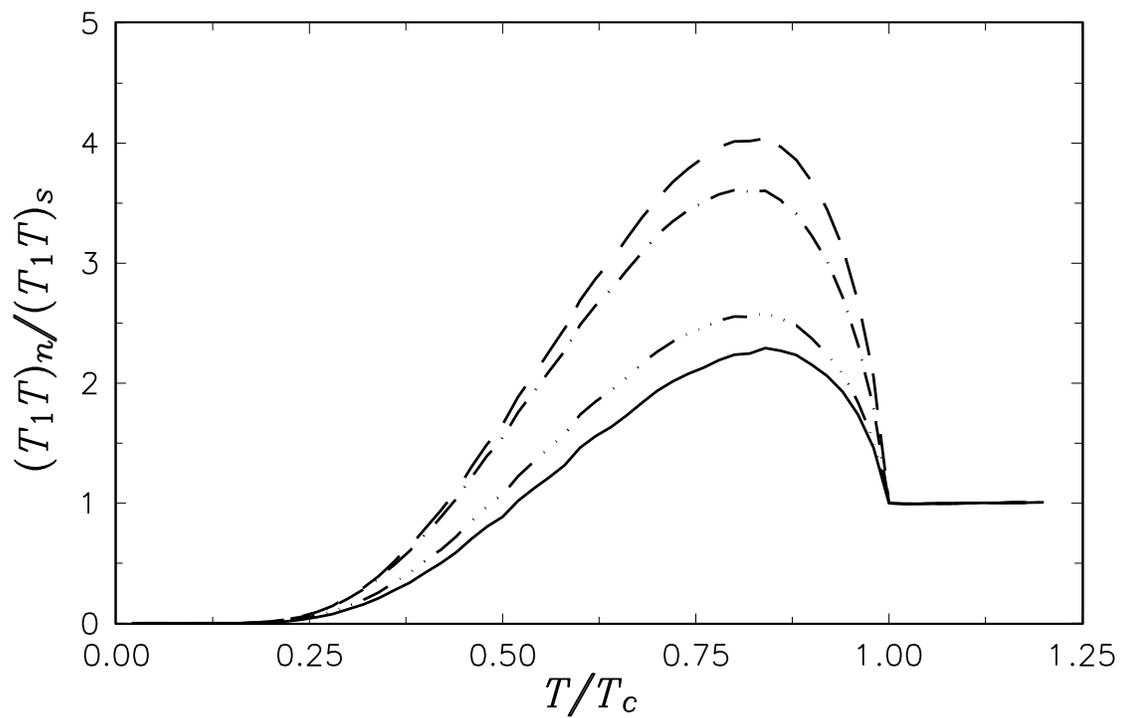
\end{document}